# Latent Pigments Strategy for Robust Active Layers in Solution-Processed, Complementary Organic Field-Effect Transistors

*Isis Maqueira-Albo,[a] Giorgio Ernesto Bonacchini,[a] Giorgio Dell'Erba,[a] Giuseppina Pace,[a] Mauro Sassi,[b] Myles Rooney,[b] Roland Resel,[c] Luca Beverina,[b]\* Mario Caironi[a]\**

[a] Center for Nano Science and Technology @PoliMi, Istituto Italiano di Tecnologia, via Pascoli 70/3, 20133 Milano (MI), Italy

[b] Department of Materials Science, University of Milano-Bicocca, Via Cozzi 53, I-20125 Milano (Italy) 70/3, 20133 Milano (MI), Italy

[c] Institute of Solid State Physics, Graz University of Technology, Petersgasse 16, 8010 Graz (Austria)

# Abstract

Solution-processed organic semiconductors enable the fabrication of large-area and flexible electronics by means of cost-effective, solution-based mass manufacturing techniques. However, for many applications an insoluble active layer can offer technological advantages in terms of robustness to processing solvents. This is particularly relevant in field-effect transistors (FET), where processing of dielectrics or barriers from solution on top of the semiconductor layer typically imposes the use of orthogonal solvents in order not to interfere with the nanometer thick accumulation channel. To this end, the use of latent pigments, highly soluble molecules which can produce insoluble films after a post-deposition thermal cleavage of solubilizing groups, is a very promising strategy. In this contribution, we demonstrate the use of tert-Butyloxycarbonyl (t-Boc) functionalized diketopyrrolopyrrole and perylene-diimide small molecules for good hole and electron transporting films. t-Boc thermal cleavage produces a densification of the films, along with a strong structural rearrangement of the deprotected molecules, strongly improving charge mobility in both p- and n-type FET. We also highlight the robustness of these highly insoluble semiconducting layers to typical and aggressive processing solvents. These results can greatly enhance the degree of freedom in the manufacturing of multi-layered organic electronic devices, offering enhanced stability to harsh processing steps.

# Introduction

Solution processable organic semiconductors have been widely studied and developed as they offer appealing opportunities in the broad field of large-area, flexible and portable electronics,[1] comprising solar-cells,[2] light-emitting devices,[3] sensors,[4,5] thermoelectrics,[6] microelectronics,[7-9] and bioelectronics.[10] The ease in tailoring the opto-electronic properties of organic semiconductors allows for the development of rich libraries of specific compounds. These compounds together with soluble conductors and dielectrics enables the fabrication of complex organic devices solely through solution-based methods. Therefore, technologies based on cost-effective, low temperature and scalable processes such as high-throughput coatings,[11-14] high resolution printing,[15-18] and direct-writing [19] have been targeted and are on the verge of producing the first real marketable applications.[20]

Since an optimized electronic device is typically composed of several functional layers that have to be deposited one on top of the other, an obvious requirement is that the deposition of a new layer must not alter the electronic properties of the underlying layers.[21-23] One of the most exploited approaches to comply with such a requirement is to rely on orthogonal solvents, i.e. solvents that solubilize the compound being deposited but cannot solubilize the layers below.[24] Obviously, in the case of multi-layer stacks, such an approach poses stringent constraints on the processing solvents. This complicates process flows and limits possible choices in terms of materials. For such a reason, future solution-based organic electronic technologies would strongly benefit from the availability of semiconducting materials that are highly soluble in common solvents and that can readily become insoluble after deposition.

Cross-linking of the active layers has been proposed as a possible solution,[25] where the need of the addition of additives may interfere with both molecular packing and charge transport.[9] In the case of polymers there are examples of efficient strategies which do not degrade the semiconductor properties, such as versatile photocrosslinking of sterically hindered

bis(fluorophenyl azide)s based on alkyl side-chain insertion reactions.[25] However, no general approaches have been developed in this direction as of yet. Additionally, in the case of small molecules the required presence of a dense, electronically inactive alkyl chain phase may not be optimal.

A promising alternative mimics the so-called latent pigment methodology, first reported by Zambounis *et al.*,[26] where insoluble pigments are turned into soluble dyes by the reversible functionalization with labile functionalities that can then be cleaved through either a thermal or optical process. This procedure unlocks molecular interactions such as hydrogen bonding and π-π interactions. Protective groups, such as the tert-butoxycarbonyl residue (t-Boc), make the derivative soluble by hindering hydrogen bonds. One of the main advantages of this approach is the possibility to remove the t-Boc radical in a post-processing step by thermal, acid or UV treatment directly after the film preparation. It is well known that t-Boc group undergoes thermolysis in quantitative yield at around 180 °C with two gaseous products, $CO_2$ and isobutene, which consequently evaporate from the solid-state film structure.

This approach can be adopted both for small molecules[27, 28] and for polymers,[29-32] and in general it could bring the additional advantage of removing the insulating solubilizing chains thus increasing the volume fraction of conjugated material in the active layer. It has previously been adopted in organic solar cells.[28, 33] The same strategy is very attractive also for the processing of robust organic field-effect transistors (FETs),[27, 32] which are typically composed of very thin, tens of nanometers films. The active semiconducting layer can be easily damaged by solvents used to deposit either barrier layers or dielectrics on top.[34] In the case of staggered, top-gate architectures, which are preferable to optimize energetics at the semiconductor-dielectric interface and charge injection,[35] the soluble dielectric is processed right on top of the active interface where the nanometer thick channel is accumulated. The use of processing solvents in this case has profound implications for the charge-transport properties, with possible creation of extended interphases.[24, 36, 37] Recently, the advantages of the latent pigment approach in the fabrication of top gate FETs was nicely demonstrated for polymers, specifically for copolymers based on isoindigo derivatives functionalized with branched chains and cleavable t-Boc groups. By post-film-casting thermal treatment, the t-Boc groups could be removed efficiently to generate a H-bond cross-linked polymer network, leading to air stable mobility values up to $10^{-4}$ cm$^2$/Vs and excellent robustness to solvents.[34]

Latent pigments can offer an even more valuable approach in the case of soluble small molecules, which offer simpler synthetic protocols, easier purification and no polidispersity with respect to polymers.[9] The choice of solvents for layers lying on top of small molecules films is severely restricted due to typically high solubility, with options relegated for example to perfluorinated solvents,[38, 39] thus posing a strong limitation on the possible dielectric materials to be adopted. So far, in the case of small molecules, the latent pigment approach has been adopted in a single case and only in a bottom-gate bottom-contact FET architecture. Mobilities, achieved for thiophenes with a diketopyrrolopyrrole core or quinacridone after the thermal cleavage of the t-Boc groups, were in the order of $10^{-6}$ cm$^2$/Vs.[27] The latent pigment approach applied to small molecules has never been employed so far neither for the development of top-gate FETs, nor for n-type FETs in general, the latter path being further complicated by the typical lower environmental stability of organic semiconductors with high electron affinity.[40-42] Examples of a high electron affinity, t-Boc functionalized small molecule was only reported for a PDI-based latent pigment used in luminescent solar concentrators.[43]

In this contribution, we report the use of thermo-cleavable t-Boc functionalized diketopyrrolopyrrole (DPP) and perylene-diimide (PDI) small molecules for the preparation of highly insoluble semiconducting layers in solution-processed top-gate FETs, thus enabling both p- and n-type staggered transistors and paving the way for the possible future use of this strategy in robust complementary circuits. The p-type FET is based on a terthiophene DPP molecule with t-Boc groups on the lactam N,N positions (T3DPP-t-Boc, Figure 1a), which has previously shown good potential for organic solar cells, but was not investigated for transistors applications.[44] For the n-type FET, we synthesized a PDI derivative with t-Boc groups in the N, N positions (PDI-t-Boc, Figure 1b). In both cases, the thin film morphology undergoes a strong rearrangement upon thermal deprotection of the t-Boc groups. This is accompanied by a drastic increase of field-effect currents, leading to promising charge mobility, up to $10^{-2}$ cm$^2$/Vs for holes transport with ambient stability. Notably, we demonstrate that the deprotected layers are so robust that rinsing of the semiconductor even with aggressive solvents still allows optimal field-effect behavior. Furthermore, the top dielectric can be processed from typically forbidden solvents, yielding good performing FETs.

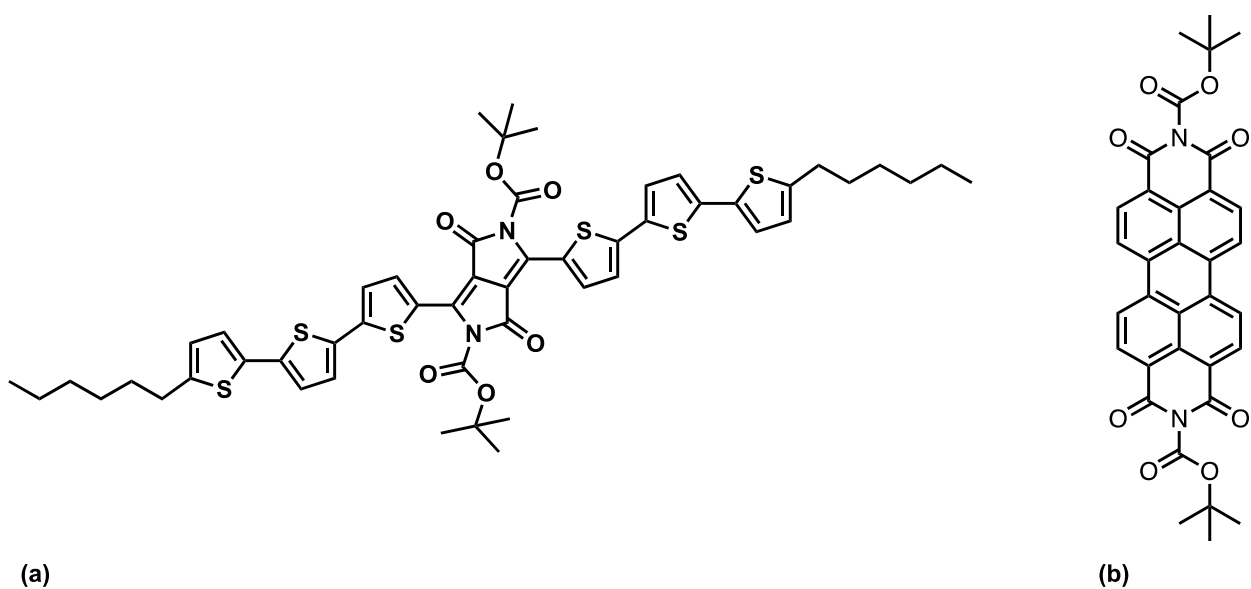

(a)  (b)

*Figure 1. Molecular structure of (a) T3DPP-t-Boc and (b) PDI-t-Boc.*

## Deprotection of molecules in the solid state

The thermal cleavage of both T3DPP-t-Boc and PDI-t-Boc was characterized by means of thermogravimetric analysis (TGA). Figure S1 of the Supporting Information shows the TGA trace of T3DPP-t-Boc. The first weight loss, corresponding to 20 % of the original mass, starts at 160 °C and is complete slightly above 180 °C. This can be attributed to the evolution of $CO_2$ and isobutene from the two t-Boc functionalities. The thus obtained T3DPP pigment remains stable up to 370 °C. Figure S2 shows the same analysis carried out on PDI-t-Boc. The thermal decomposition of the t-Boc groups happens in this case at the significantly lower temperature of 110-130 °C. The PDI pigment remains stable up to 550 °C.

Specular X-ray scans over a thermal gradient were conducted on drop cast films (see Supporting Information, Figure S3), identifying the beginning of structural changes at around 160-170 °C, which are completed within 10 minutes at 170 °C. These processes occur more rapidly at 200 °C, requiring only a few minutes, as evidenced by the UV-Vis experiments depicted in Figure S4. The absorption spectra were recorded after annealing times of the T3DPP-t-Boc films ranging from 30 seconds to 10 min. After 2 min, there is no evidence of further changes in the absorptions bands of the film.

On the basis of the previous evidences, we set the deprotection temperature at 200 °C, applied for 5 minutes (to ensure a complete cleavage of the t-Boc groups). We then studied its effect on thin films of both compounds deposited by spin coating from a chloroform solution, first by means of UV-Vis absorption (Figure 2). In the case of the T3DPP-t-Boc film, before deprotection (Figure 2a, red line) the features observed in the absorption spectrum are in line with previous studies on similar molecules with a DPP core,[33, 45] according to which the absorption bands around 400 nm can be attributed to the oligothiophene component of the molecule, the peak at roughly 660 nm is attributed to a charge transfer state associated with the DPP core while the peak at 735 nm relates to solid-state aggregation effects.[45] Upon the thermal treatment at 200 °C the absorption bands are strongly blue shifted with respect to the non-deprotected films, with the appearance of a sharp feature at 563 nm that can be attributed to a densification effect after the removal of the t-Boc groups, accompanied with an improved packing of the molecules and stronger π-interactions of the aromatic rings, giving rise to the formation of intermolecular H-aggregates.[33]

The PDI-t-Boc film prior to deprotection shows an absorption maximum peak at around 550 nm together with two lower intensity vibronic peaks at around 507 and 470 nm corresponding to the strongly allowed $S_0$-$S_1$ transition of PDI with the transition dipole moment oriented along the PDI molecules axes, in agreement with previous results. [25] After the thermal treatment, a redistribution of the oscillator strength between the two main bands (550 and 507 nm) occurs. The higher energy band is associated with an H-type molecular stacking, while the lower energy band corresponds to a J-type arrangement. [46-50] Therefore, the increased intensity of the absorption band at around 507 nm after the deprotection, constitutes an indication of the formation of prevalent H-type aggregation.

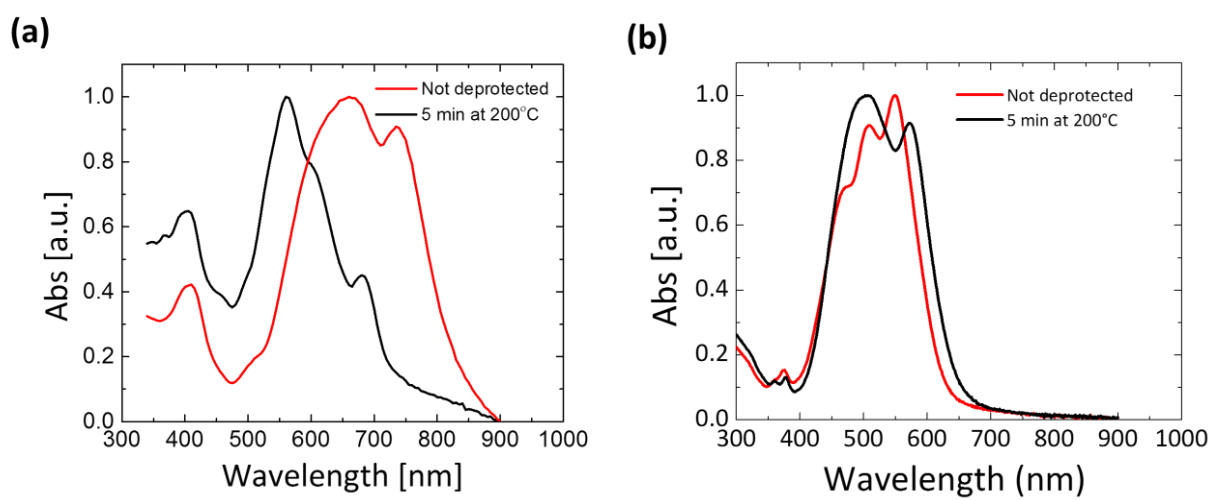

*Figure 2.* UV-Vis characterization of T3DPP-t-Boc (a) and PDI-t-Boc (b) thin films.

Overall, UV-Vis absorption suggests strong rearrangements within the thin films following chemical deprotection, with enhanced molecular interactions. We investigated the underlying structural changes within the films with X-Ray Diffraction (XRD), X-Ray Reflectivity (XRR) and Atomic Force Microscopy (AFM). From XRR measurements (Figures S5 and S6) of spin cast T3DPP-t-Boc films, there is a significant decrease in film thickness from 40 to 29 nm and an increase in surface roughness from 1.3 to 3 nm. An increase in the critical angle of the organic material from 0.17° to 0.18° indicates an increased electron density within the film from 435 $nm^{-3}$ before cleavage to 488 $nm^{-3}$ after cleavage.

The AFM topography of the T3DPP-t-Boc film before deprotection (Figure 3b) is characterized by rod-like beads, the dimension of which appear smaller upon deprotection (Figure 3c). The corresponding XRD patterns (Figure 3a), measured on the same film, show a diffraction peak at $2\theta = 7.7°$ (d = 11.5 Å) before deprotection. This peak corresponds to the second order of the peak observed in the X-Ray Reflectivity measurements (Figure S3) at $2\theta = 3.8°$ and constitutes an indication of the presence of a crystalline phase. Such peak completely disappears with cleavage, after which only a small and broad signal at low angle values is present.

A similar morphological modification can be observed in PDI films (Figure 3e, f). Topography shows more spherical domains in this case, which have smaller dimensions after deprotection. Correspondingly, the diffraction peak at $2\theta = 4.5°$ (d = 19.6 Å) present in the PDI-t-Boc films (Figure 3d) is no longer present for the deprotected ones, as observed also in the case of the T3DPP-t-Boc films. The presence of a Bragg peak before deprotection (at $2\theta = 7.7°$ and 4.5° for T3DPP-t-Boc and PDI-t-Boc respectively) originates from an out-of-plane molecular packing motif with respect to the substrate. These peaks disappear after deprotection, following the strong structural rearrangement, with indication of a possible re-orientation of the molecules with respect to the substrate plane, and/or to a very strong reduction of the crystallites sizes.

Overall, the films upon deprotection appear to be more densely packed, as demonstrated by XRR, with improved molecular interactions owing to the removal of the solubilizing groups and thus to an increased number of tight contact points between $\pi$ orbitals of molecules. Such modifications upon cleavage should favor intermolecular charge transfer and lead to an easier establishment of an efficient charge transport pathway.

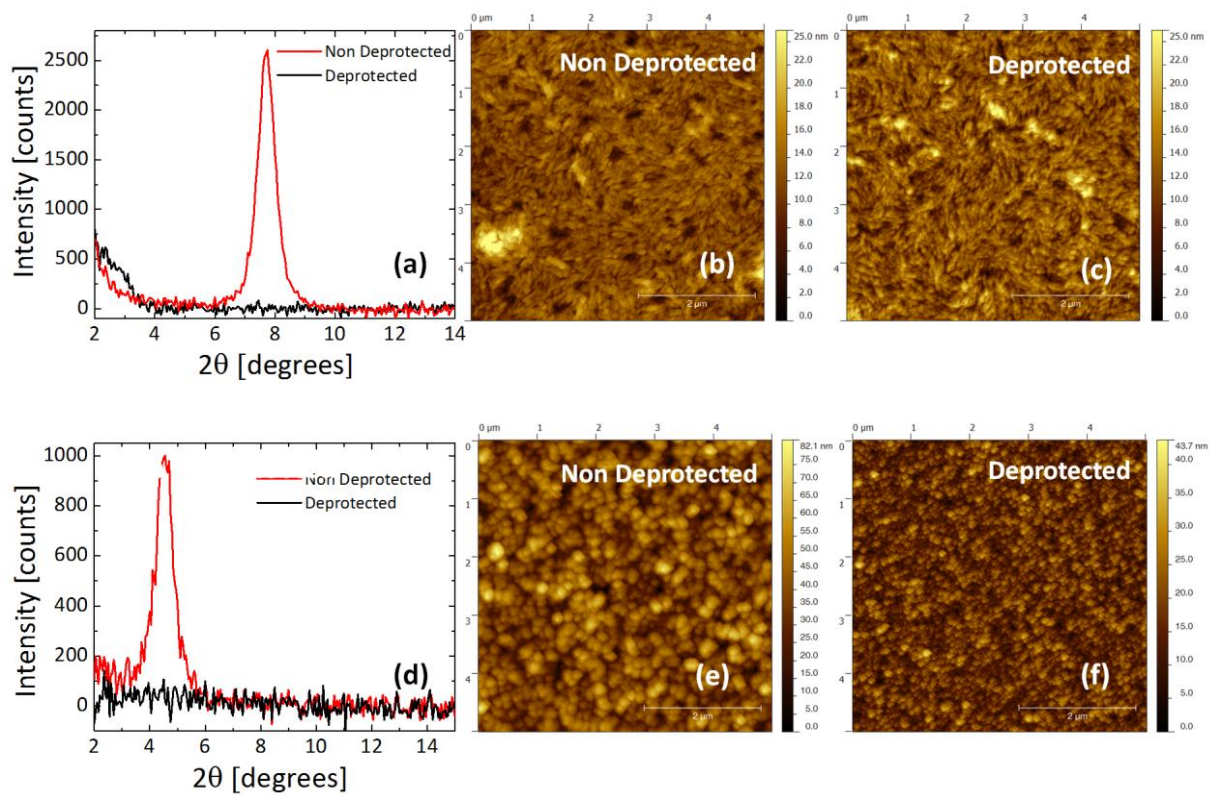

*Figure 3.* XRD and AFM topography images for T3DPP-t-Boc (a,b,c) and PDI-t-Boc (d, e, f) films before and after deprotection.

**Field-effect transistors**

We tested the charge transport properties of T3DPP and PDI films before and after deprotection in staggered, top gate FETs. The semiconducting films were deposited on the interdigitated gold electrodes following the same spin-coating process adopted for the samples characterized above. At first, we adopted a typical orthogonal solvent, n-butyl acetate, widely used in top-gate solution-processed organic FETs, to deposit the top PMMA dielectric. The devices based on T3DPP that did not undergo thermal cleavage of the solubilizing group show a strongly unbalanced ambipolar behavior, in favor of a hole accumulation regime (Figure 4 a and Figure S7). When measured in p-type regime, typical transfer curves can be observed, with $10^4$ on-off ratio in linear regime and signature of an n-type accumulation at high source-drain voltage ($V_{SD}$) and gate-source voltage ($V_{GS}$) around 0 V. Upon thermal treatment at 200 °C in air, the hole current increases by more than an order of magnitude. The n-type channel is completely suppressed, making the device substantially unipolar (Figure 4a, Figure 4c and Figure S8) in linear and saturation regimes, with on-off ratios in the order of $10^5$. We found that the suppression of the electron conducting channel strictly depends on the deprotection atmosphere. As shown in Figure S9 in the Supporting Information, electron accumulation and transport is still possible upon deprotection in a nitrogen atmosphere, leading in any case to a strongly unbalanced ambipolarity. Although a clear-cut explanation of the phenomenon requires specific studies, such behavior could be explained by considering that the cleaved molecules N-H functionality could react with $CO_2$, thus leading to a lowering of the HOMO and LUMO levels (due to the electron-withdrawing nature of the carbonyl group) as well as to silencing of the hydrogen bond interactions. Obviously, when the cleavage is performed under nitrogen, this process is suppressed. This kind of $CO_2$ capture effect has already been described for quinacridone, a pigment structurally similar to DPP.[51] Output curves show that even though $N_2$-deprotected devices show increased hole mobility with respect to the air-deprotected ones (factor 3), slightly better injection is achieved for the latter (Figure 4c).

The observed variations in the currents upon deprotection are reflected in the extracted FET mobility, which for holes increases in saturation from $2\times10^{-4}$ cm$^2$/Vs before deprotection, to 0.01 cm$^2$/Vs and 0.03 cm$^2$/Vs when deprotection is performed in air and nitrogen, respectively. Electron mobility is $5\times10^{-5}$ cm$^2$/Vs prior to the thermal treatment, and it does not substantially vary with deprotection in nitrogen ($4\times10^{-5}$ cm$^2$/Vs).

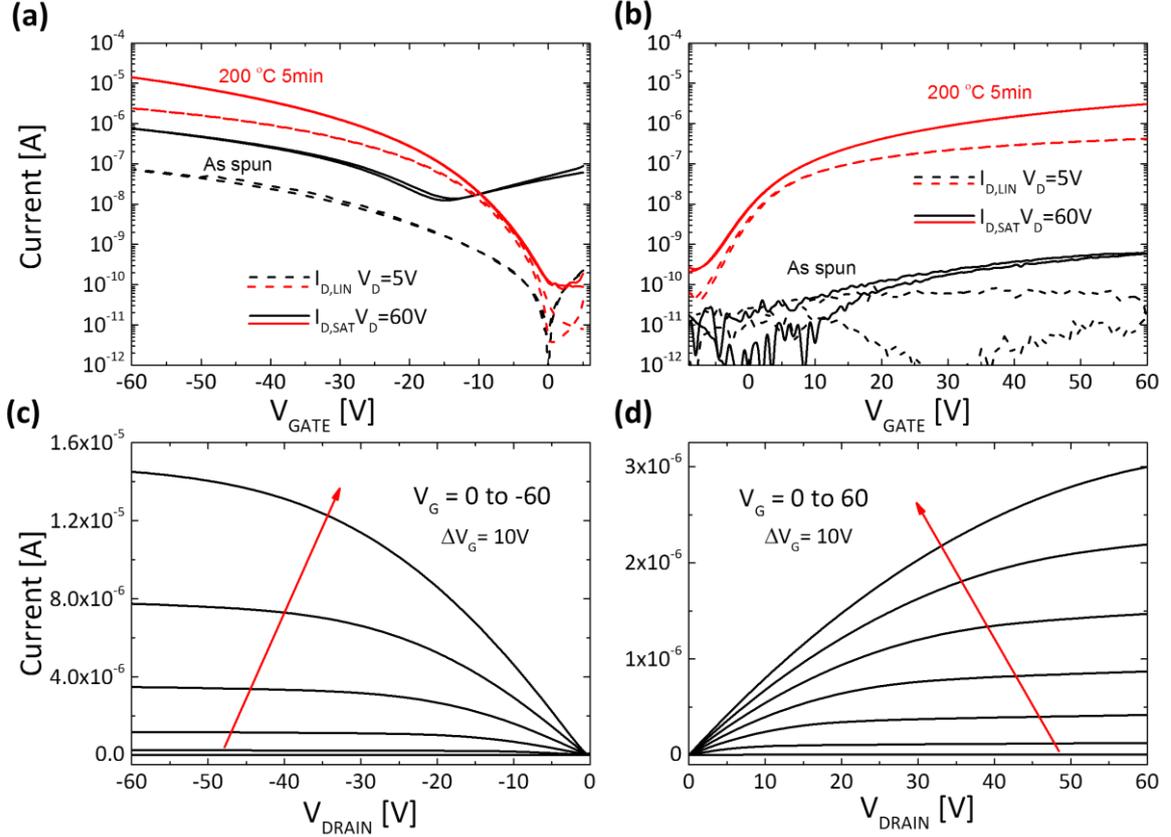

*Figure 4. Electrical characterization of latent pigments based FETs. Transfer curves for (a) T3DPP before and after deprotection, (b) PDI before and after deprotection. Output curves for (c) T3DPP deprotected (d) PDI deprotected. All the devices had W = 10000 µm and L = 20 µm.*

In the case of the PDI films, very weak electron current modulation is present for the non deprotected molecules, with an estimated electron mobility around $10^{-7}$ cm$^2$/Vs in the saturation regime, with no measurable hole current (Figure 4b and Figure 4d). The device performance drastically improves when the semiconductor film is deprotected in air, with an increase of more than 3 orders of magnitude in the currents and a corresponding increase in electron mobility, reaching 4x$10^{-4}$ cm$^2$/Vs in saturation (Figure 4b). Despite the limited value of charge mobility recorded in this case, this is the first example of the use of the latent pigment approach for n-type FET devices.

Mobility values for all FETs here studied are summarized in Table 1. A remarkable improvement of the transport properties for both T3DPP and PDI based FETs is present after the cleavage of the t-Boc solubilizing group, which is accompanied by the strong reorganization of the films previously described. Previous works on solution-processed DPP or indigo-based pigments generally show maximum carrier mobilities in the range of $10^{-2}$ cm$^2$/Vs [45, 52, 53]. However, reports in which the latent pigment approach for DPP based molecules is studied, report lower mobility values with respect to our work ($10^{-6}$ cm$^2$/Vs for DPP-t-Boc [27] and $10^{-3}$ cm$^2$/Vs DPP based co-oligomers [54]). Similarly to our observations, in the case of DPP co-oligomers, performances of the devices were improved after the thermal cleavage of t-Boc groups, where charge carrier mobilities were enhanced by two orders of magnitude, going from $10^{-5}$ to $10^{-3}$ cm$^2$/Vs. This improvement was attributed to the formation of H-bonded networks.[54] In our work, the effect is more pronounced owing to the small size of molecules. As such, the weight fraction of insulating material (t-Boc) removed by thermal treatment is higher with respect to bigger molecular systems. Therefore, the removal of the solubilizing groups results in denser films with stronger intermolecular interactions, favoring the formation of efficient charge percolation paths throughout the FET channels.

*Table 1. Charge mobility extracted from T3DPP and PDI based FETs.*

|  | T3DPP | | PDI |
|---|---|---|---|
| Device | $\mu_h$ [cm$^2$/Vs] | $\mu_e$ [cm$^2$/Vs] | $\mu_e$ [cm$^2$/Vs] |
| Non Deprotected | 2 x 10$^{-4}$ | 4 x 10$^{-5}$ | 10$^{-7}$ |
| Air Deprotected | 0.01 | - | 4 x 10$^{-4}$ |
| N$_2$ Deprotected | 0.03 | 4 x 10$^{-5}$ | - |

**Stability tests**

Since the presented FETs undergo a thermo-chemical reaction within their active phase, it is critical to verify the operational stability of the devices under electrical stress. We performed gate bias stress tests in dark conditions on both p- and n-type FETs with deprotected materials in nitrogen and in air. The most demanding stability test for FETs is the gate bias stress test, where the device is constantly kept at a constant $V_{GS}$ voltage, and therefore with a constantly formed channel. Typically, with time, a threshold voltage $V_{TH}$ shift ($\Delta V_{TH}$) is observed due to trapping of accumulated carriers in non-mobile, deep trap states related to chemical instability, intrinsic structural defects (e.g., grain boundaries), and/or external impurities and/or chemical functionalities present at the dielectric interface.[55, 56] This phenomenon can be electrically monitored by recording the channel current $I_{DS}$ in time upon constant biasing (fixed $V_{DS}$ and $V_{GS}$). Here we have recorded $I_{DS}$ versus time curves under $|V_{GS}| = 40$ V and $|V_{DS}| = 40$ V for 2 hours in nitrogen, and then for 2 additional hours in air. Transfer curves were measured in nitrogen at the beginning of the test, in air before and after the stress in nitrogen and after the total 4 h test. This protocol was designed to be able in principle to decouple intrinsic stress effects, eventually induced in thermally cleaved films with respect to air induced effects.

Figure 5a highlights the current behavior during the overall stress measurement for T3DPP based OFETs. During the bias stress test performed in nitrogen atmosphere, a slow decrease in current over time is observed. After 2 h, a 30 % current loss is registered, reaching 102 nA, corresponding to a slight $\Delta V_{TH}$ of 0.48 V (Figure 5c). Another bias stress test was carried out on a different T3DPP device in air (blue dashed line in Figure 5a). After 2 h only a 20 % current loss is registered, surprisingly displaying a lower current loss rate than in inert atmosphere. We then took the device previously stressed in nitrogen and subjected it to a bias stress test in air. Starting from $I_{DS} = 110$ nA, owing to a partial device recovery after the first stress, the current increases with time, and it shows a 20 % current gain in 2 h. The cumulative effect of the two biass stress tests – first under nitrogen and then under air - is 13 % loss of the initial current. The behavior of the device under air exposure is due to a doping of the semiconductor through an electrically triggered oxidation effect. This is evident in the final transfer curves which show much higher, un-gateable off currents, which strongly reduce the on-off ratio to $10^1$ (Figure 5c). The doping likely occurs through the whole film thickness, with respect to a field-effect modulation which is effective only on the first few molecular layers at the interface with the dielectric, originating the high off conductivity.[57]

Similar measurements were performed for the PDI based OFETs (Figure 5b,d). The current reduction registered during the bias stress test in nitrogen accounts only to 19 % of the initial current over 2 h. Another 10 % of current reduction occurs as soon as the device is exposed to air, with $\Delta V_{TH}$ = 4.06 V (Figure 5d).[58] The subsequent bias stress in air indicates a good ambient stability for an n-type device, with only a further 15 % current reduction (Figure 5b). Transfer curves measured during the stress test confirms the good stability, with a $\Delta V_{TH}$ of 3.02 V.

In the case of T3DPP, the electrically activated doping effect effectively compensates for the reduction in operating current when the device is directly measured in air (around 20 % current loss), limiting the reduction of the current driving capability of our device to around 13 % of the initial value at the conclusion of the 2 h stress test. It is the on-off ratio which is not stable upon gate bias stress in air, but this is a limit of the conjugated core of the molecule, characterized by a not low enough HOMO energy level, and it is not due to the latent pigment approach. It is therefore expectable that the adoption of the same strategy with conjugated cores characterized by a lower HOMO level would retain as well the on-off ratio.

The air stability of our n-type OFETs is remarkably superior to previous reports in literature based on PDI, as this small molecule operates poorly in air unless it is provided with fluorocarbon substituents at the N,N' positions or with electron-withdrawing groups such as CN or F. Very interestingly, the stability measurements of our devices are comparable with gate bias stress tests performed on cyano-substituted PDI, probably owing to the dense packing of the small molecules as well as the encapsulating effect of the PMMA insulating layer, both slowing the permeation of $O_2$ and $H_2O$ within the semiconducting film.[40, 41, 59]

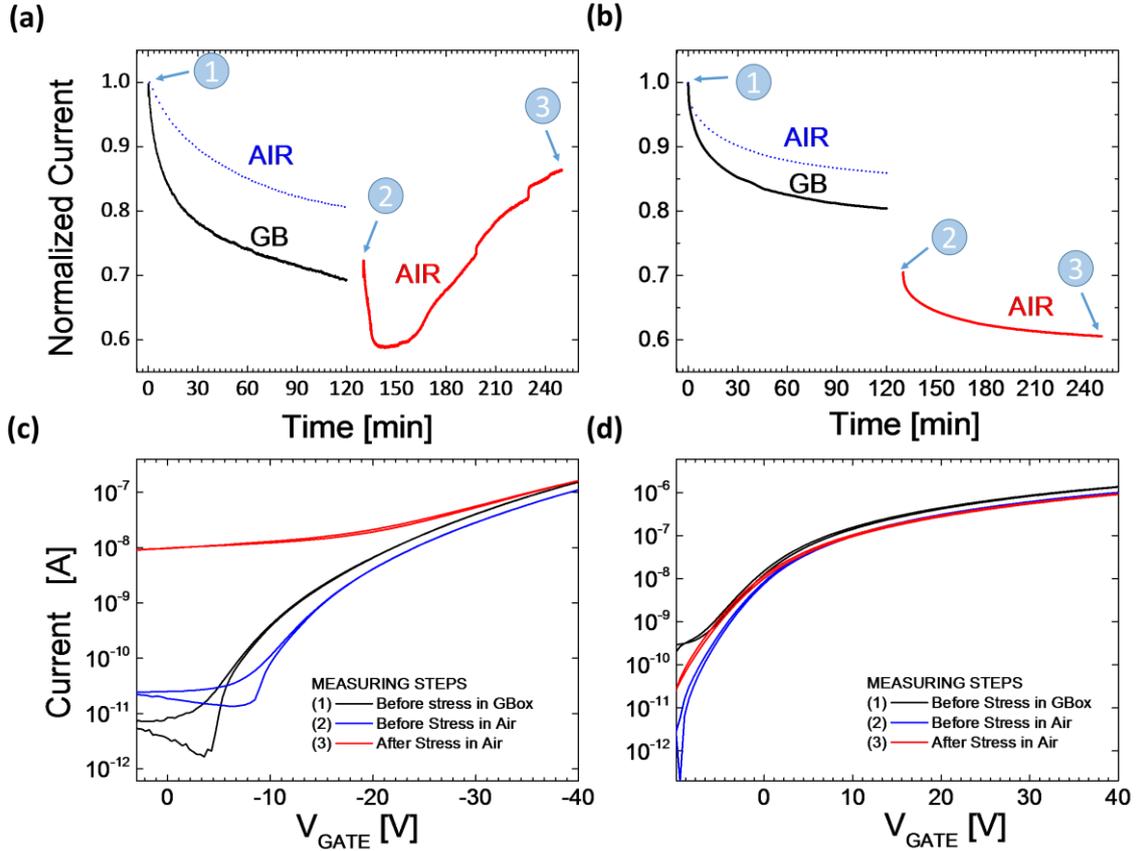

**Figure 5.** *Sequential stress-test on a T3DPP and PDI device under nitrogen (solid black line) and then in air (solid red line). Blue dashed line corresponds to a different device stress-test directly in air. $I_{DS}$ is recorded every second for 2 hours at $|V_{GS}|=|V_{DS}|=40$ V both in nitrogen atmosphere and in air for (a) T3DPP and (b) PDI. Transistor transfer curves for T3DPP (c) and PDI (d), taken before stress tests in inert atmosphere (black) and in air (blue), and after the stress test carried in air (red). All the devices had W = 10000 μm and L = 20 μm.*

**Robustness to solvents**

To prove the effectiveness of the adopted strategy in yielding electronic devices that, upon thermal deprotection, are robust with respect to processing solvents, we performed two different tests on the T3DPP semiconductor: *i)* rinsing of the semiconductor film with various solvents before deposition of PMMA from nBA, and *ii)* depositing the PMMA layer from different solvents, typically not allowed because of strong interaction with most organic semiconductors.

We want to underline that for a top-gate small molecule FET these are particularly harsh conditions, as we directly expose the surface of the semiconductor to aggressive solvents. This region hosts the charge accumulation layer, the thickness of which varies with the density-of-states of the specific semiconductor, and it is in the few nanometers range.[57]

Robustness to rinsing was assessed by spinning chloroform (CF), chlorobenzene (CB) and N,N-Dimethylformamide (DMF) directly onto the semiconductor surface, followed by 10 min annealing at 200 °C when CB or DMF were deposited. The device was then completed by depositing the dielectric and the top-gate to measure the effect of the solvent exposure. Results of this test are shown in Figure 6a. All devices maintain a good FET behavior, similar to the reference device that was not exposed to solvents. In the case of the low boiling point CF, no substantial difference can be seen, while $\Delta V_{TH} \approx 4$ V can be observed when CB or DMF, which have higher boiling points, are used.

We then fabricated T3DPP FETs by adopting CF and CB as solvents for PMMA, two solvents that readily dissolve the protected T3DPP molecule. These devices were then compared to reference devices made with n-BA as the dielectric solvent. Data shown in Figure 6b demonstrates that it is possible to fabricate good working FETs using typically forbidden solvents for the deposition of the dielectric. Mild $\Delta V_{TH}$ of 1.48 V for CF and 2.28 V for CB and saturation currents of $1.06 \times 10^{-5}$ A and $5.6 \times 10^{-6}$ A are measured, respectively. Although there may not be a strong reason to choose CF or CB for processing PMMA, this proof-of-concept experiment shows that the latent pigment strategy is effective in widening the choice of processing solvents for multi-layer devices and the library of functional materials that can be stacked one on top of the other from solution.

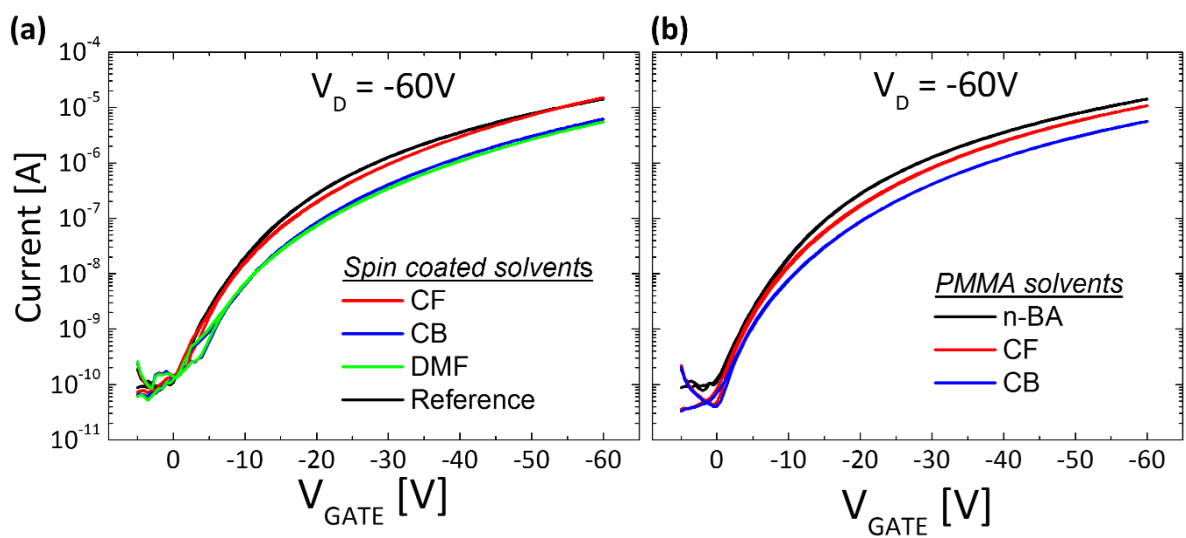

*Figure 6. Robustness test of T3DPP based OFETs. (a) semiconductor film after deprotection with direct solvent spin-coating and (b) by changing the dielectric solvent. All the devices had W = 10000 µm and L = 20 µm.*

## Conclusions

A latent pigment approach has been employed for the fabrication of both p- and n-type solution-processed organic FETs. t-Boc functionalized T3DPP and PDI molecules can be readily dissolved in common organic solvents and can be therefore deposited from solution to form thin semiconducting films. A thermal process performed in ambient air can successively be adopted to deprotect the molecules in the solid-state, triggering a strong structural rearrangement leading to thinner, more densely packed films with prevalent H-type aggregation. Electrical measurements on FETs based on both molecules show that field modulated currents are strongly increased when the molecules are deprotected. In the case of T3DPP an enhancement of two orders of magnitude is recorded for holes and three orders of magnitude for electrons in PDI. This indicates that the thermal cleavage and enhanced intermolecular interactions following the films structural rearrangement contribute to the formation of efficient conducting channels for charge. P-type FET based on deprotected T3DPP show a hole mobility of ~$10^{-2}$ cm$^2$/Vs, with unipolarization of the device when t-Boc cleavage is performed in air. These values are one order of magnitude higher with respect to previous reports on DPP based co-oligomers [54] and four orders compared to DPP-t-Boc small molecules.[27] The realization of the n-type latent pigment FETs is reported, to the best of our knowledge, for the first time in our work in the case of PDI, showing good FET characteristics and electron mobility in the order of ~$10^{-4}$ cm$^2$/Vs upon ambient air deprotection. Promising stability to gate bias stress has been recorded in nitrogen and ambient air. This is particularly relevant for the electron transporting material. As a proof-of-concept, we tested the robustness of the deprotected semiconductor to harsh solvent treatments, which would typically wash away the soluble small molecules semiconducting layer. The solvent treatments did not prevent the observation of good field-effect behavior, demonstrating that the high insolubility of the films can preserve the materials good electrical properties. This enables the use of typically forbidden solvents for the deposition of functional layers on top of the semiconductor. While charge mobility is still clearly limited with respect to state-of-the-art solution processed organic semiconductors, we have demonstrated that the protection of organic semiconductors with t-Boc, their processing from solution and the subsequent cleavage and removal of these solubilizing moieties, is a feasible strategy to develop highly insoluble and robust films with both hole and electron transporting properties. In particular, we have unlocked the use of this strategy for small molecules, with field-effect mobility four orders of magnitude higher than previous attempts, and for top-gate devices, thus critically expanding its applicability in many contexts. Our results therefore pave the way for the adoption of the latent-pigment strategy into a solution-based multi-layer processing technology. We envision this as a useful method to be integrated into large scale fabrications such as printing complementary organic circuit applications and other devices requiring heterojunctions such as solar cells and detectors. The accurate control of final microstructure by tuned chemical design and material optimization coupled with investigation of further cleavable functional groups will offer the possibility to improve device performances to meet requirements for such applications.

## Experimental

**Synthesis of the materials**

Compound T3DPP-t-Boc was prepared by a slight modification of the protocol which we previously reported (see Figure S10 Supporting Information for the details).[33] PDI-t-Boc was prepared according to the reported procedure by Mattiello *et al.* [43]

**Films preparation**

T3DPP-t-Boc and PDI-t-Boc were dissolved in chloroform (5 g/l) and deposited by spin-coating at 1000 rpm for 60 s in air on low-alkali Corning 1737F glass. For the cleavage of the t-Boc group the films were annealed at 200 °C in air. In the case of T3DPP-t-Boc the cleavage of the t-Boc in inert atmosphere was also tested. The final film thickness was about 50 nm.

The T3DPP-t-Boc films tested for x-ray reflectivity were prepared by spin-coating 8mg/ml in chloroform at 4000 rpm for 40 seconds onto cleaned silicon dioxide 1x1cm substrates. The samples were made in tandem with one half being treated thermally in an oven at 200°C for 7 mins. The second half underwent a thermal gradient experiment with an inbuilt heating plate in the XRD instrument.

**Films characterization**

The TGA characterization of latent pigments was performed on traces acquired on a Thermogravimetric Analyzer Mettler-Toledo (TGA/DSC) at a heat scan ratio of 5 °C/min under a nitrogen flux of 50 ml/min.

The surface topography of the films was measured with an Agilent 5500 Atomic Force Microscope operated in the Tapping Mode. The UV-Vis characterization of thin films was performed on a Varian's Cary 50 UV-Vis Spectrophotometer. X-ray diffraction analysis was performed with a Bruker AXS 2009 diffractometer with a Cu Kα source (1.54 Å wavelength) and the background was corrected by subtracting the X-ray spectra of the glass substrate to the X-ray spectra of the material under study.

The X-ray reflectivity measurements were carried out on an PANalytical Empyrean system with a 1/32° divergence slit, a 10 mm beam mask, a multilayer x-ray mirror for monochromatisation and parallelizing of the x-ray beam and an attenuator at the primary side, radiation from a sealed copper source (CuKα) was used. At the secondary side a 0.1 anti-scatter slit, a 0.02 radian Soller slit and a PANalytical PIXCEL detector was used. The experimental data were fit with the software X'Pert Reflectivity which used Parrat formalism[60] using the approach of Nevot & Croice[61] for the determination of surface roughness.

**FET fabrication**

A Top-Gate, Bottom-Contact architecture was employed in the FETs fabrication. The Bottom contacts were prepared by a lift-off photolithographic, where a 1.5 nm thick Cr adhesion layer followed by a 50 nm Au contact were deposited by thermal evaporation. The channel width ($W$) and length ($L$) are 10000 μm and 20 μm, respectively. Prior to the semiconductor deposition the patterned substrates were rinsed with acetone and isopropanol. After the deposition of the semiconductor, a 600 nm thick layer of PMMA (Sigma-Aldrich, $M_w$ = 120 kg/mol) was spun from n-butyl acetate (with a concentration of 80 g/l) at 1250 rpm. For the robustness tests of the semiconductor layer the PMMA was deposited also from chloroform and chlorobenzene at a concentration of 80g/l and spun at 700 and 1300 rpm respectively. After the dielectric deposition, the devices were annealed at 80°C for 30 min in inert atmosphere. Subsequently, 40 nm thick aluminum gate electrodes were thermally evaporated as gate contacts.

**FET electrical measurements**

The measurements of the transfer and output characteristics of the FET devices were performed in a nitrogen glove box using an Agilent B1500A Semiconductor Parameter Analyzer. Mobility extraction was carried out at $V_G = V_D = 60$ V by plotting $\mu_{SAT}$ vs $V_G$, according to:

$$\mu_{SAT} = \frac{2L}{C_{diel}W}\left(\frac{\partial\sqrt{I_D}}{\partial V_G}\right)^2 \qquad (1)$$

Where $C_{diel}$ is the gate dielectric capacitance per unit area, $\mu_{SAT}$ is the carrier mobility in the semiconductor, $I_D$ the drain current, $V_G$ and $V_D$ the gate and drain voltage respectively.

**Supporting Information**

Supporting Information is available from the Wiley Online Library or from the author.

**Acknowledgements**

This work has been in part financially supported by the European Research Council (ERC) under the European Union's Horizon 2020 research and innovation programme 'HEROIC', grant agreement 638059. I. M.-A. acknowledges support of the European Union through the Marie Curie Initial Training Network POCAONTAS (FP7-PEOPLE-2012-ITN No. 316633). L. B. and M. R. acknowledge support of the European Union through the Marie Curie Initial Training Network Thinface (FP7-PEOPLE-2013-ITN No. 607232).

Supporting Information

**Latent Pigment Strategy for Robust Active Layers in Solution-Processed, Complementary Organic Field-Effect Transistors**

*Isis Maqueira-Albo,[a] Giorgio Ernesto Bonacchini,[a] Giorgio Dell'Erba,[a] Giuseppina Pace,[a]*

*Mauro Sassi,[b] Myles Rooney,[b] Roland Resel,[c] Luca Beverina,[b*] Mario Caironi[a*]*

[a] Center for Nano Science and Technology @PoliMi, Istituto Italiano di Tecnologia, via Pascoli 70/3, 20133 Milano (MI), Italy

[b] Department of Materials Science, University of Milano-Bicocca, Via Cozzi 53, I-20125 Milano (Italy) 70/3, 20133 Milano (MI), Italy

[c] Institute of Solid State Physics, Graz University of Technology, Petersgasse 16, 8010 Graz (Austria)

*E-mail of corresponding author: mario.caironi@iit.it, luca.beverina@mater.unimib.it

# TGA CHARACTERIZATION OF LATENT PIGMENTS

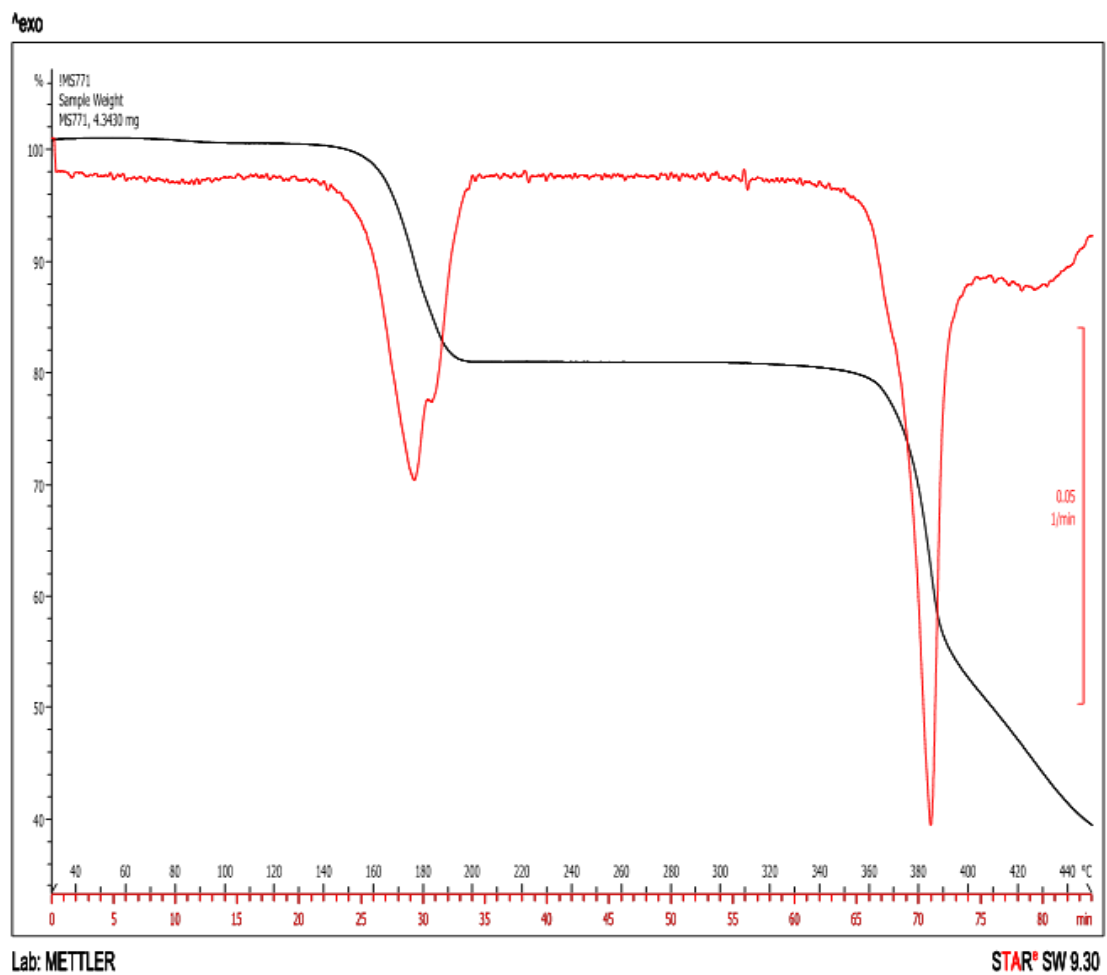

*Figure S1. TGA (black trace) and DSC (red trace) of a T3DPP-t-Boc sample.*

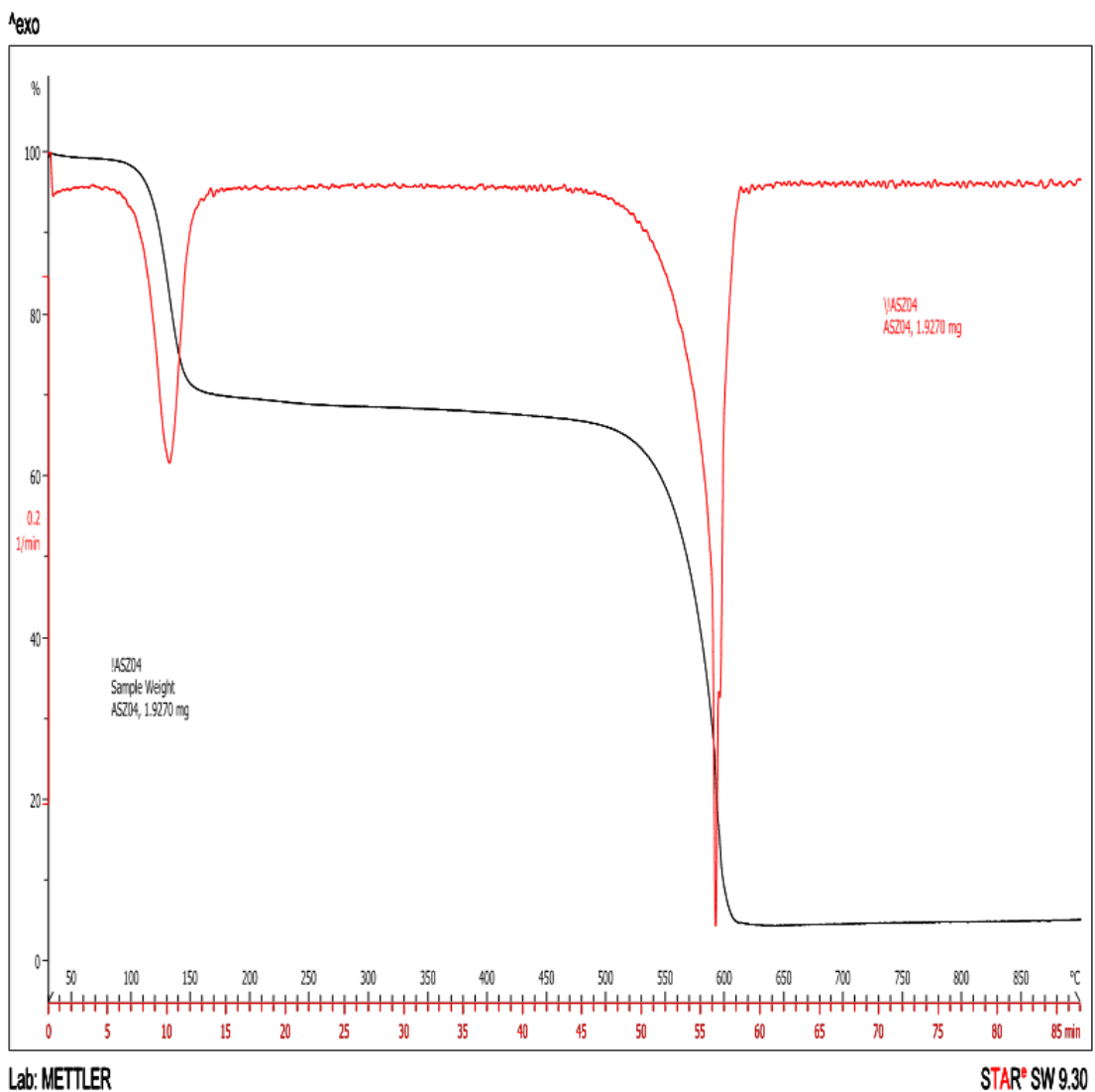

*Figure S2. TGA (black trace) and DSC (red trace) of a PDI-t-Boc sample.*

## X-RAY AND UV-VIS MEASUREMENTS

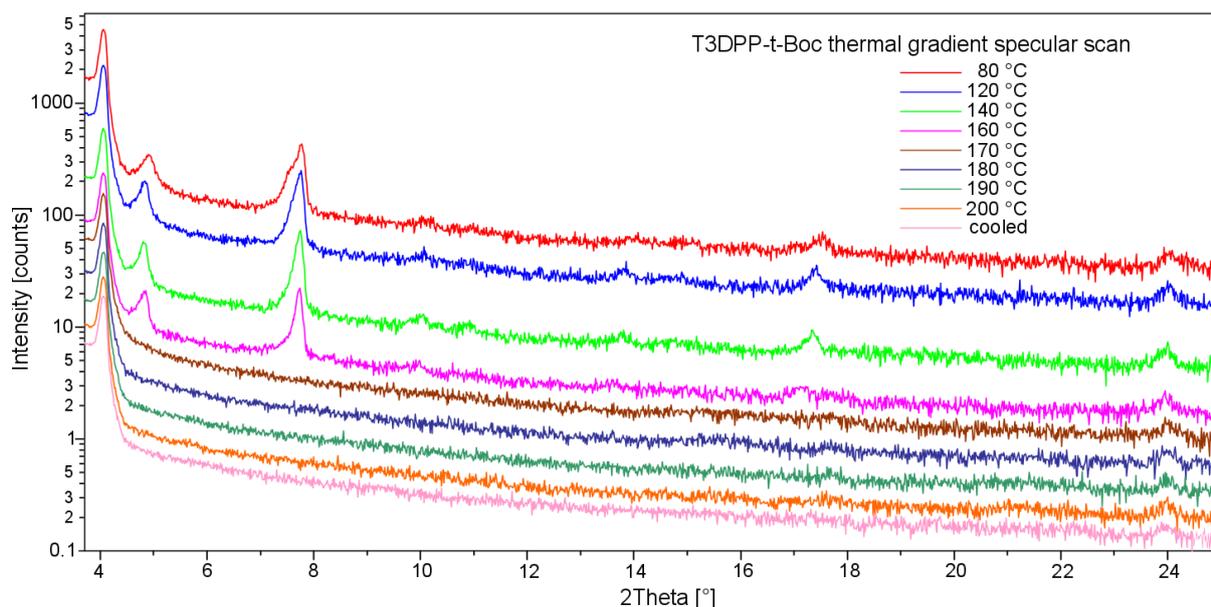

*Figure S3: Thermal gradient specular scan of a T3DPP-t-Boc drop-cast film from CHCl$_3$ in Air*

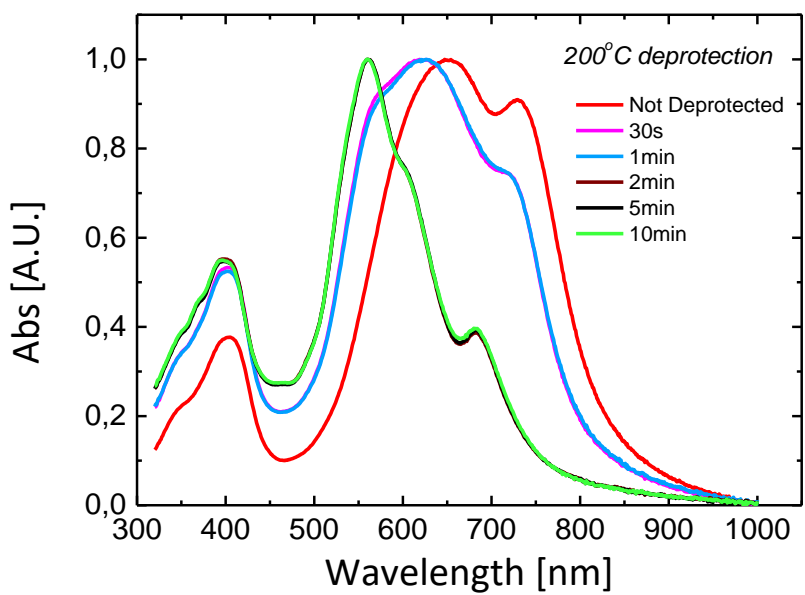

*Figure S4: UV-Vis characterization of T3DPP-t-Boc thin films at different times of annealing at 200 °C. After 2 min the complete cleavage of the t-Boc group is achieved.*

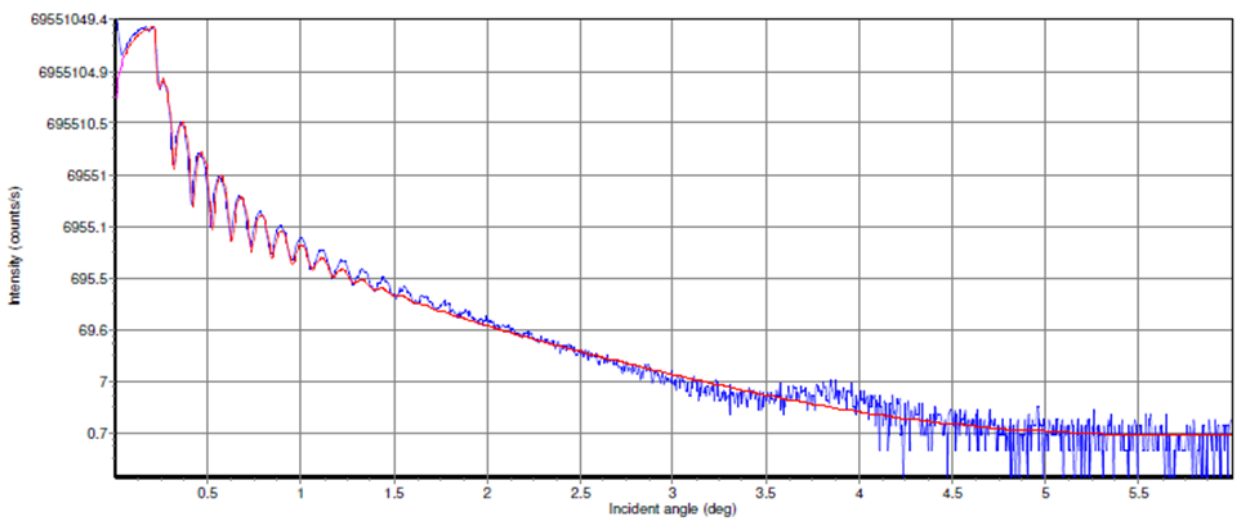

*Figure S5: X-Ray Reflection measurement of a spin coated T3DPP-t-Boc film*

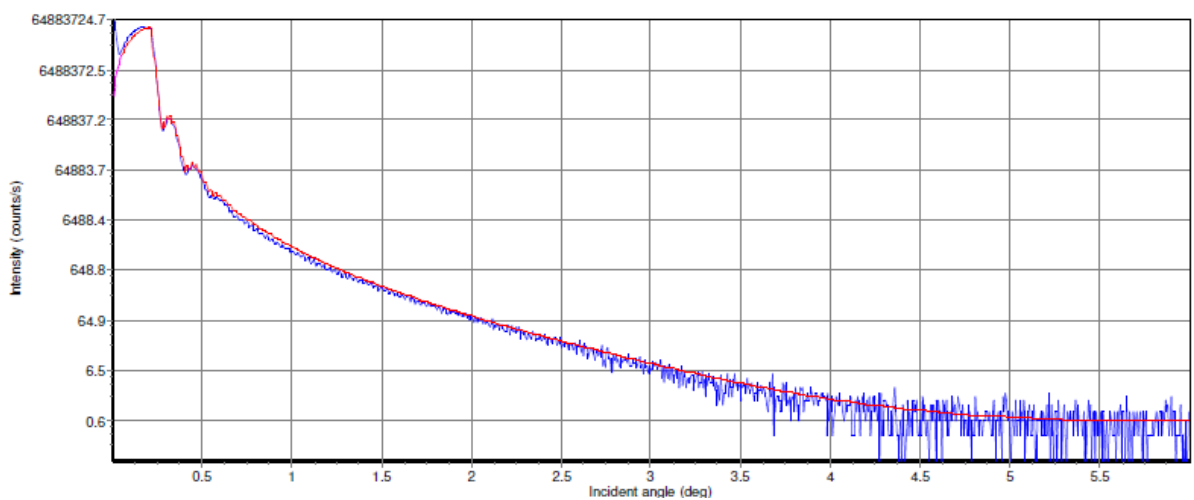

*Figure S6: X-Ray Reflection measurement after thermal cleavage of the t-Boc group of a spin coated T3DPP-t-Boc film*

There is a significant decrease in film thickness upon thermal deprotection, from 40 to 29 nm, and an increase in surface roughness from 1.3 to 3 nm. The film thickness after thermal cleavage (72 % of its initial value) can be almost directly related to the loss of mass within the film as the tert-butyloxycarbonyl groups are removed:

$$\frac{797.17\ g/mol}{997.4 g/mol} \times \frac{100}{1} = 79\%\ of\ mass\ remaining\ after\ deprotection$$

$$\frac{29nm}{40nm} \times \frac{100}{1} = 72\%\ of\ film\ thickness$$

*Table S1.* *Summary of the T3DPP-t-Boc film characteristics before and after the thermal cleavage of the t-Boc group extracted from the X-Ray Reflection measurements.*

| Characteristics | T3DPP-t-Boc | T3DPP-t-Boc annealed |
|---|---|---|
| SiO$_2$ critical angle | 0.225 | 0.2252 |
| organic layer critical angle | 0.17 | 0.18 |
| Organic electron density (electrons per nm$^3$) | 435 | 488 |
| Roughness (nm) | 1.3 | 3 |
| Thickness (nm) | 40 | 29 |
| d-spacing (of Bragg peak) (nm) | 1.14 (typical of edge on alignment) | No Bragg peak seen |
| Vertical crystal size (nm) | 13 | N/A |

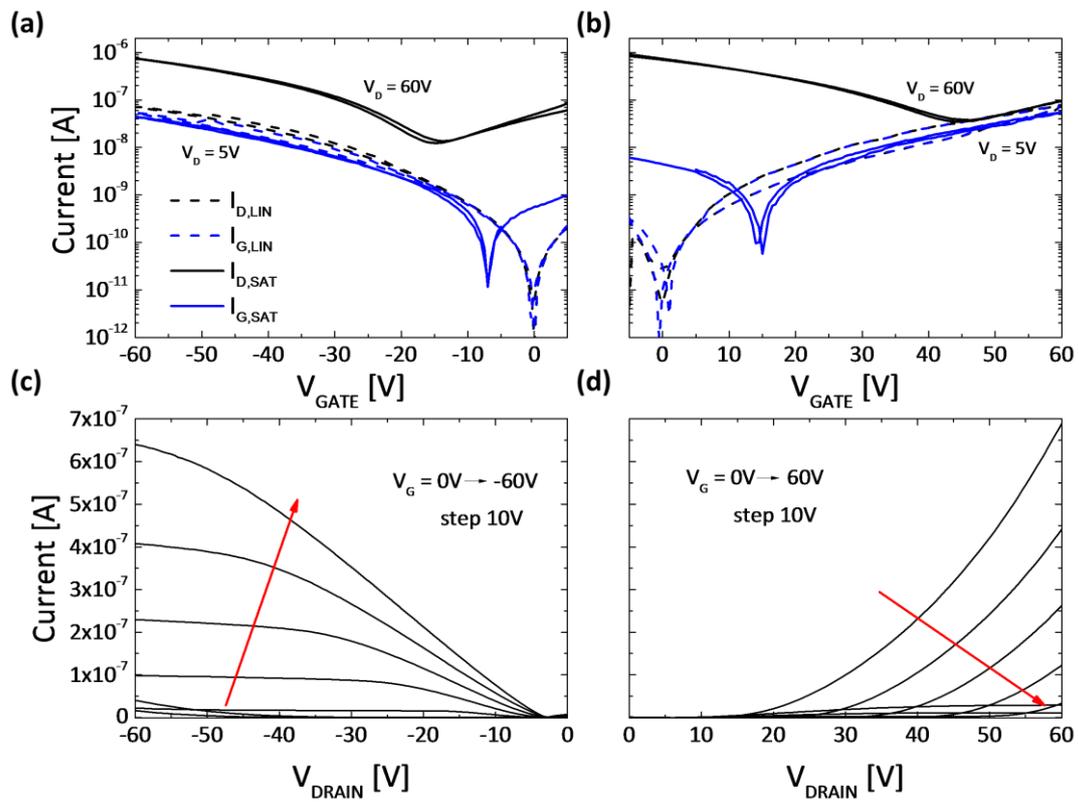

*Figure S7: Electrical characterization of T3DPP-t-Boc FETs non deprotected. . Transfer curves for hole (a) and electron (b) conduction of a T3DPP-t-Boc device. Output curves for hole (c) and electron (d) conduction. All the devices had W = 10000 μm and L = 20 μm.*

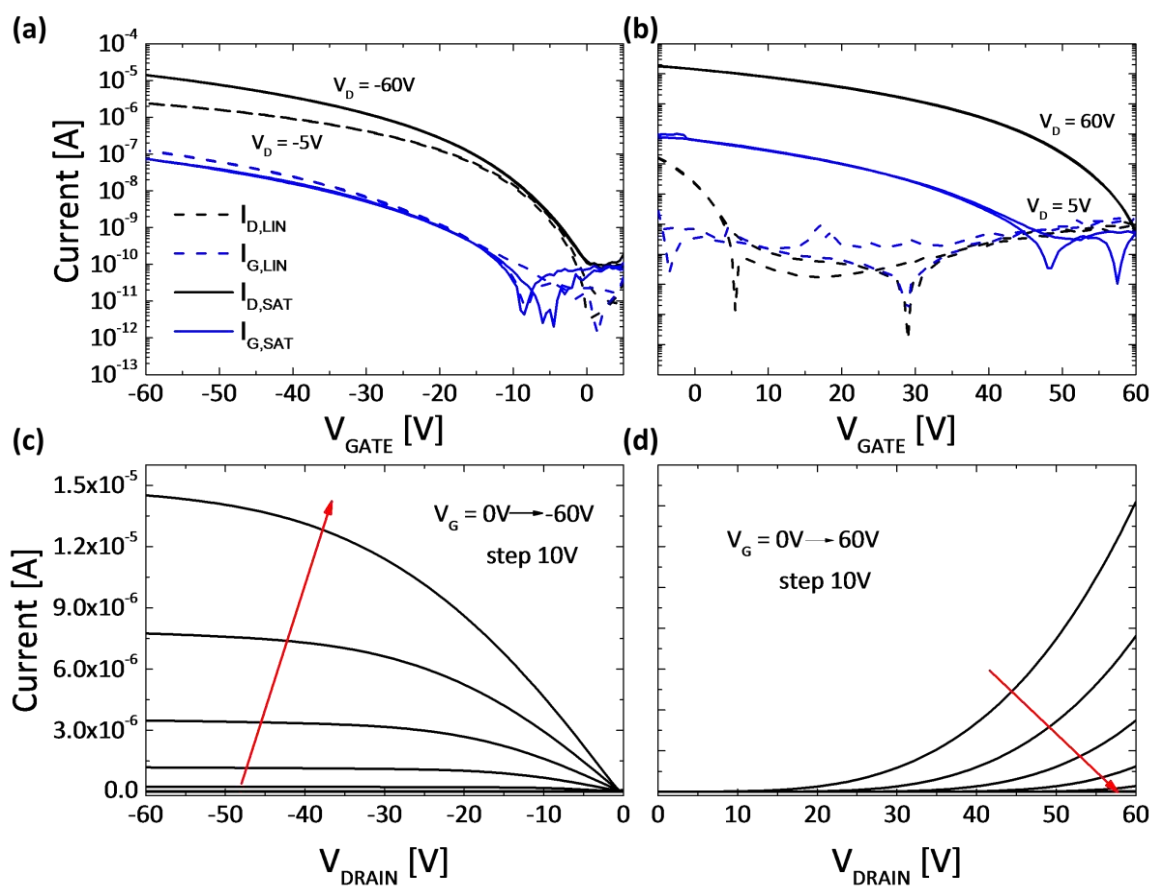

*Figure S8: Electrical characterization of T3DPP FETs deprotected in air. Transfer curves for hole (a) and electron (b) conduction of a T3DPP device. Output curves for hole (c) and electron (d) conduction. All the devices had W = 10000 μm and L = 20 μm.*

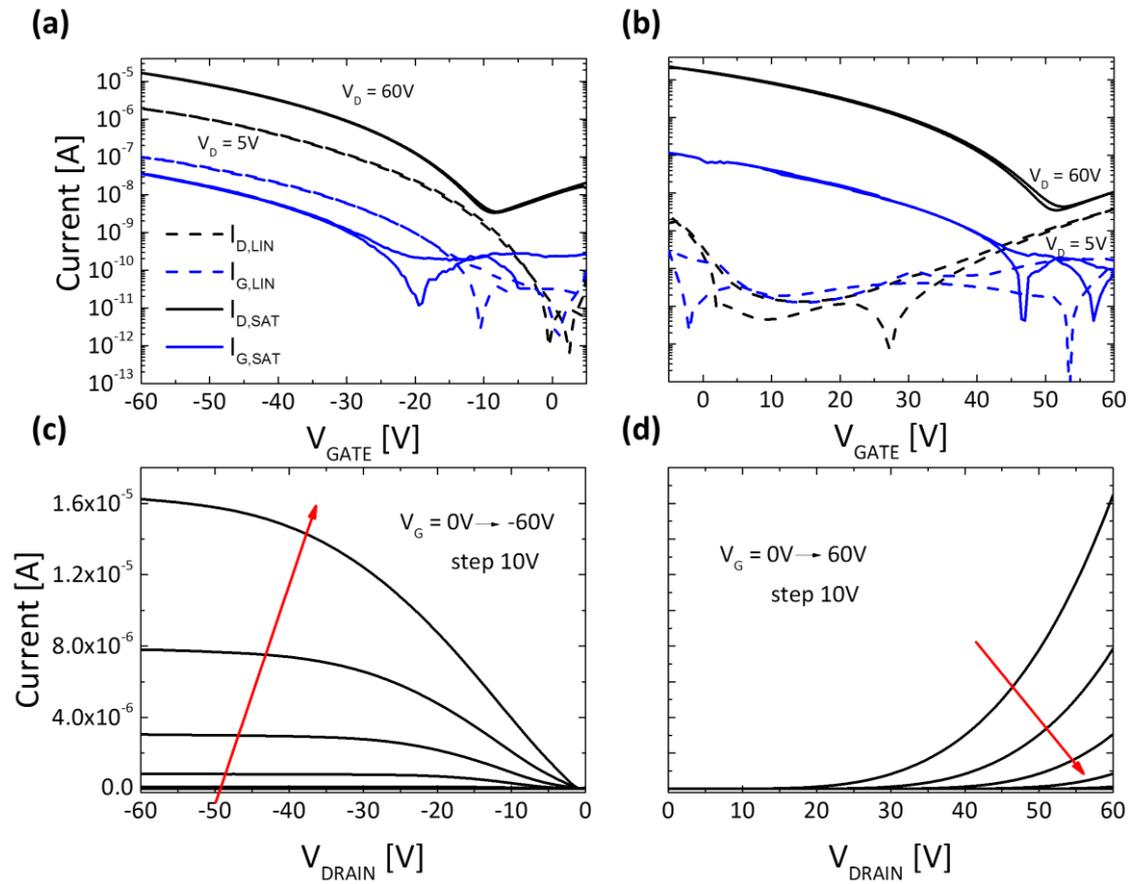

*Figure S9: Electrical characterization of T3DPP FETs deprotected under nitrogen atmosphere. Transfer curves for hole (a) and electron (b) conduction of a T3DPP device. Output curves for hole (c) and electron (d) conduction. All the devices had W = 10000 µm and L = 20 µm.*

**SYNTHESIS OF T3DPP-T-BOC**

1. 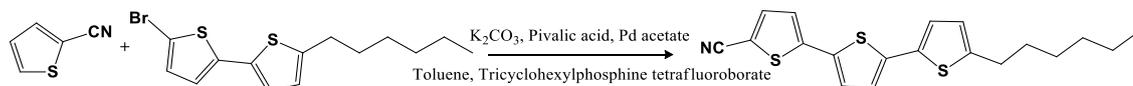

2. 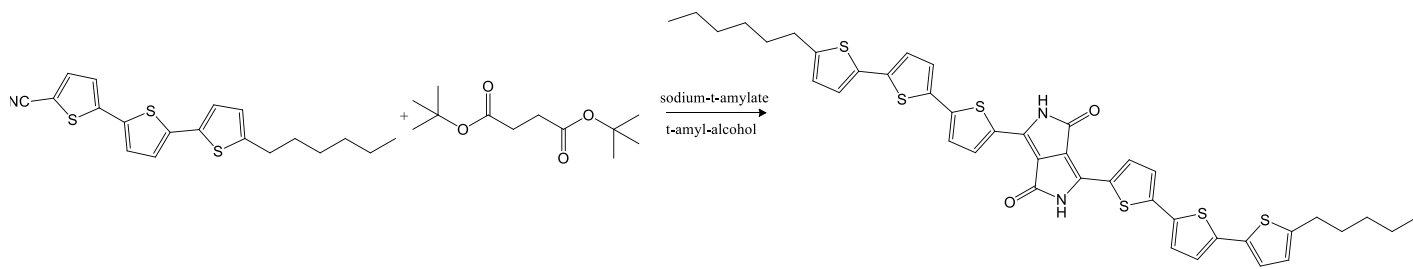

3. 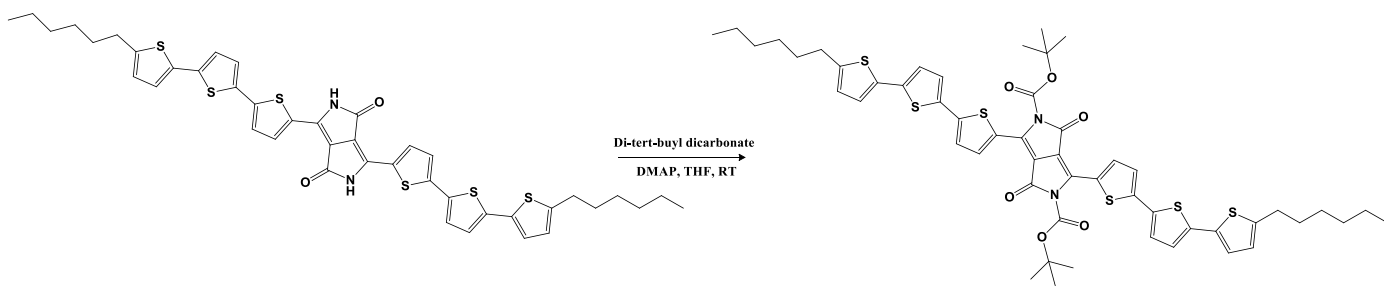

*Figure S10: T3DPP-t-Boc molecule synthesis scheme*

**1.** In a Nitrogen Glove box, charge 3g (9.1mmol) 5-bromo-5'hexyl-2,2'bithiophene, 2.615g (23.96mmol) thiophene-2-carbonitrile, 2.483 (17.97mmol) $K_2CO_3$, 53.79mg(0.24mmol) Palladium acetate, 195mg (0.529mmol) tricyclohexylphosphine tetrafluoroborate, 367mg(3.6mmol) Pivalic acid and 40ml dry toluene into a 100ml pressure tube with magnetic stirrer. Seal cap tightly. The solution has a light yellow colour. Heat to 150 °C for 24 hours. Allow to cool to room temperature before pouring solution into 100ml water. Dilute with 100ml toluene and separate. Wash organic phase with 2x100ml water and 100 ml brine. Dry organic phase on $MgSO_4$. Remove solvent with rotary Evaporator. A dark brown/yellow oil is recovered. Purify with column chromatography 3:7 toluene: cyclohexane gradient to 1:1 toluene: cyclohexane. Recover a yellow solid. The final yield of this step is 40%. Characterization data are in agreement with those previously reported. [1]

**2.** In a 50ml tri-neck round bottom flask with Dean stark and dropping funnel dissolve 0.52g(4.7mmol) sodium-t-amylate in 5ml tert-amyl-alcohol under nitrogen atmosphere. In a separate flask under nitrogen dissolve 1.3g(3.6mmol) 5"hexyl-[2,2':5',2"-terthiophene]-5-carbonitrilein 5ml dry THF before transferring to the 50ml RBF. Heat this solution to 120°C to distill off the THF with magnetic stirring. Into the dropping funnel add dropwise over 90minutes a solution of tert-butyl-succinate 0.364g(1.58mmol) dissolved in 12ml THF. The solution should darken immediately upon addition. Stir for 3 hours at 120°C. Cool to 50°C add 1.5ml isopropyl alcohol and stir for a further 10minutes before adding 2ml distilled water. Take up solution in 10ml isopropyl alcohol and filter through fluted filter paper. Was extensively with 50:50 boiling iPrOH:$H_2O$. Stir the crude powder in boiling THF and Hot filter. Dry in Vacuum oven to afford 750mg of very dark powder. The final yield of this step is 52%. The product was not further purified and used directly in the next step. NMR was not taken due to lack of solubility.

**3.** In a 50ml round bottom flask with magnetic stirrer charge 0.75g (0.95mmol) of 3,6-bis(5"-hexyl-[2,2':5',2"terthiophene]-5-yl)pyrrolo [3,4-c]pyrrole-1,4(2H,5H)-dione and 0.29g(2.37mmol) of Dimethylaminopyridine under Nitrogen atmosphere. Dissolve in 10ml Dry THF. In a separate flask under Nitrogen dissolve 1.5g(6.8mmol) in 5ml Dry THF and transfer to the main reaction flask at room temperature. Stir vigorously for 18hrs.solution darkens with green particles dispersed. Filter and wash with 20ml THF followed by 20ml cold diethyl ether and 50ml 1:1 diethyl ether: Petroleum ether to afford dark green powder. The final yield of this step is about the 90%. 1H NMR (CDCl3) $\delta$: 8.28 (2H, d, J=4.20 Hz), 7.20 (4H, dd, J=4.20, 3.86 Hz), 7.03 (4H, dd, J=3.75, 3.45 Hz), 6.70 (2H, d, J=3.55 Hz), 2.80 (4H, t, J=7.61 Hz), 1.68 (4H, quint, J=7.45 Hz), 1.42-1.28 (12 H, m), 0.9 (6H, t, J=7.01 Hz). 13C

NMR (CDCl3) d: 159.97, 149.88, 147.42, 144.88, 140.16, 137.33, 136.34, 134.87, 134.84, 128.66, 127.17, 125.93, 125.09, 124.97, 124.77, 111.10, 86.88, 32.43, 32.41, 31.10, 29.62, 28.64, 23.44, 14.96. Anal Calcd for $C_{52}H_{56}N_2O_6S_6$: C, 62.62; H, 5.66; N, 2.81. Found: C, 62.65; H, 5.59; N, 2.78.